\documentclass[universe,article,submit,moreauthors,pdftex,10pt,a4paper]{mdpi} 
\firstpage{1} 
\articlenumber{0000}
\doinum{10.3390/------}
\pubvolume{xx}
\pubyear{2016}
\copyrightyear{2016}
\externaleditor{Academic Editor: name}
\history{Received: date ; Accepted: date ; Published: date}

\usepackage{amssymb,latexsym}
\usepackage{tensor}
 \theoremstyle{mdpi}
 \newcounter{thm}
 \newcounter{ex}
 \newcounter{re}
\newcommand{\mc}{\mathcal}

\newcommand{\bs}{\begin{split}}
\newcommand{\es}{\end{split}}

\newcommand{\eqr}{\eqref}
\newcommand{\g}{g\indices}
\newcommand{\Q}{\mathcal{Q}}
\newcommand{\A}{\mathcal{A}}
\newcommand{\dr}{\ell}

 \theoremstyle{mdpidefinition}

\Title{On the Near Horizon Canonical Quantum Microstates from $AdS_2/CFT_1$ and Conformal Weyl Gravity}

\Author{Leo Rodriguez $^{1*}$ and Shanshan Rodriguez $^{2}$}

\address{%
$^{1}$ \quad Department of Physics, Assumption College, Worcester, MA 01609; ll.rodriguez@assumption.edu\\
$^{2}$ \quad Department of Physics, Worcester Polytechnic Institute, Worcester, MA 01605; slrodriguez@wpi.edu}

\corres{Correspondence: ll.rodriguez@assumption.edu}

\abstract{We compute the full asymptotic symmetry group of black holes belonging to the same equivalence class of solutions within the Conformal Weyl Gravity formalism. We do this within an $AdS_2/CFT_1$ correspondence and by performing a Robinson-Wilczek two dimensional reduction, thus enabling the construction of an effective quantum theory of the remaining field content. The resulting energy momentum tensors generate asymptotic Virasoro algebras, to $s$-wave, with calculable central extensions. These centers in conjunction with their proper regularized lowest Virasoro eigen-modes yield the Bekenstein-Hawking black hole entropy via the statistical Cardy formula. We also analyze quantum holomorphic fluxes of the dual CFTs in the near horizon, giving rise to finite Hawking temperatures weighted by the central charges of the respective black hole spacetimes. We conclude with a discussion and outlook for future work.}



\begin{document}



\setcounter{section}{0} 
\section{Introduction}
Since the seminal work of Brown and Henneaux \cite{brownhenau} and the $Kerr/CFT$ correspondence \cite{kerrcft,Compere:2012jk}, it has become universally accepted that most black holes exhibit a holographic dual description in terms of a conformal field theory ($CFT$) of lesser dimensions. This duality stems in most part from the fact that the spacetime metric of the (near-)extremal black holes exhibits an $AdS$ factor in the near horizon regime. This is not the case for regular black holes which exhibit Rindler geometry in their respective near horizons and therefore makes any construction of a pure Schwarzschild/CFT or non-extremal/CFT correspondence cumbersome and still an \emph{open question} in the field of black hole physics. 

Black hole thermodynamic quantities \cite{hawk2,hawk3,beken},
\begin{align}
\label{eq:htaen}
\begin{cases}
T_H=\frac{\hbar\kappa}{2\pi}&\text{Hawking Temperature}\\
S_{BH}=\frac{A}{4\hbar G}&\text{Bekenstein-Hawking Entropy}
\end{cases},
\end{align}
are the usual tests for the above mentioned dualities. It is expected that any candidate theory of quantum gravity should contain a variant of \eqr{eq:htaen}. Since Hawking's seminal investigation of the density of quantum states in terms of Bogolyubov coefficients, the effective action approach and associated energy-momentum tensors for semiclassical matter fields have been explored in various settings for arriving at $T_H$ \cite{mukwipf,LLRphd}. Of particular interest to our analysis in this article, are the conclusion by: 
\begin{itemize}
\item Christensen and Fulling \cite{chrisfull}, whose analysis of an anomalous energy momentum tensor to compute $T_H$ was first carried out by considering the most general solution to the conservation equation
\begin{align}
\label{eq:stresscons}
\nabla_\mu T\indices{^\mu_\nu}=0.
\end{align}
They found that by restricting to the $r-t$ plane of a free scalar field in Schwarzschild geometry the energy momentum tensor exhibits a trace anomaly leading to the result:
\begin{align}
\label{}
\langle T\indices{^r_t}\rangle=\frac{1}{768\pi G^2 M^2}=\frac{\pi}{12}{T_H}^2,
\end{align}
which is exactly the luminosity (Hawking flux, Hawking radiation) of the 4-dimensional black hole in units $\hbar=1$.
\item Robinson and Wilczek (RW) who showed that anomalous two dimensional chiral theories in the near horizon of black holes are rendered unitary by requiring the black hole to radiate at temperature $T_H$ \cite{robwill,isowill,Zampeli:2012tv,srv}. RW's analysis also provides a way for obtaining near-horizon two dimensional analog black holes coupled to matter fields of parent four dimensional pure gravity solutions \cite{ry}.
\end{itemize}
 
Black hole CFT dualities rely on the conjecture that gravity in anti-de Sitter ($AdS$) space is dual to a conformal field theory \cite{Maldacena:1997re}. This paradigm has surged a tremendous program led by Strominger \cite{strom2,kerrcft}, Carlip \cite{Carlip:2011ax,carlip,carlip3,carlip2}, Park \cite{Park:1999tj,Park:2001zn,kkp} and others, in applying $CFT$ techniques to compute Bekenstein-Hawking entropy of various black holes, in particluar (near-)extremal ones. By far, the most notable example is the Kerr/$CFT$ correspondence and its extensions \cite{kerrcft,Compere:2012jk}, where the general idea is that the asymptotic symmetry group (ASG), preserving certain metric boundary or fall off conditions, is generated by a Virasoro algebra with a calculable central extension:
\begin{align}
\label{eq:vir}
\left[\Q_m,\Q_n\right]=(m-n)\Q_{m+n}+\frac{c}{12}m\left(m^2-1\right)\delta_{m+n,0},
\end{align}
where $m,n\in\mathbb{Z}$. The Bekenstein-Hawking entropy is then obtained from Cardy's formula \cite{cardy2,cardy1} in terms of the central charge $c$ and the 
normalized lowest eigenmode $\Q_0$ (without Casimir shift):
\begin{align}
\label{eq:cf}
S=2\pi\sqrt{\frac{c\cdot\Q_0}{6}}.
\end{align}
Since surface gravity is usually employed in regulating the quantum charges of \eqr{eq:vir}, thus leading to a finite $\Q_0$, there is some difficulty in using \eqr{eq:cf} for the extremal geometries where surface gravity, and therewith Hawking temperature, vanishes identically. To circumvent this,  a thermal Cardy formula is used: 
\begin{align}
\label{eq:tcf}
S=\frac{\pi^2}{3}\left(c_LT_L+c_RT_R\right)
\end{align}
where the subscripts $L$ and $R$ refer to the dual two chiral CFT theories with central charges $c_L$ and $c_R$, Frolov-Thorne vacuum temperatures $T_L$ and $T_R$ and where $c_L=c_R$ assuming diffeomorphism invariance of the two chiral halves \cite{Kraus:2005zm,Compere:2012jk,Castro:2010fd,cft}. In the specific extremal Kerr case with angular momentum per unit mass $J$ the assumption for the central charge is $c_L=c_R=c=12J$, which yields the standard entropy area law and coincides with the general expression \cite{Guneratne:2016kib,Ropotenko:2015egm,Button:2013rfa,Park:2001zn,Button:2010kg,ry}
\begin{align}\label{eq:nexctlrintro}
c=\frac{3A}{2\pi G}
\end{align}
for the central charge in terms of the black hole horizon area $A$ and Newton's constant. 

It is expected that quantum gravity (a fully formulated ultraviolet complete theory of gravity) in two dimensions should be dual to a quantum conformal field theory of equal dimensions \cite{strom2,Strominger:1994tn}, however two dimensional dilation gravity theories derived via dimensional reduction of the four dimensional Schwarzschild spacetime are in general not conformally invariant. Yet, from the $c$-Theorem \cite{carlip,ctzjet} we know that these two dimensional theories must flow, under their renormalization group, to a CFT. The consensus is that this near horizon CFT should take the form of a Liouville type theory \cite{ry,carlip3,carlip,Carlip:2001kk,deAlwis:1992hv,solodukhin:1998tc}. We also note, that the computation of either black hole temperature or entropy is traditionally addressed in separate scenarios (as described above). A novel aspect of our present calculation is the demonstration that both of these thermodynamics quantities are contained in one formalism, via a near horizon Liouville theory.

In addition to past explorations into the topic of a possible non-extremal black hole or Schwarzschild/CFT correspondence \cite{cadss,solodukhin:1998tc,Carlip:2011ax,ChangYoung:2012kd}, there seems to be renewed contemporary interest on this topic \cite{Shajiee2016,Sadeghi:2015uak,Ropotenko:2015egm} with varying strategies, approaches and advantages. The more contemporary ideas \cite{Shajiee2016,Sadeghi:2015uak} implement a conformal map which endows the Schwarzschild spacetime in four dimensions with an $AdS_2$ factor. This followed by a dimensional reduction to two dimensions within the Einstein-Hilbert action yields interesting boundary dynamics, which includes a calculable central extension. This extension together with the thermal Cardy formula reproduces the standard area law for the Schwarzschild entropy. However, the computation of the full ASG with a proper $SL(2,\mathbb{R})$ subgroup and the computation of the lowest Virasoro eigen mode, is missing and is part of the new results presented in this manuscript. In addition, since the result was derived from a Weyl rescaling of the Schwarzschild metric, it begs the question if this scenario is actually a true Schwarzschild/CFT correspondence, since the resulting rescaled spacetime is no longer of the same equivalence class as the Schwarzschild spacetime within the Einstein Hilbert action formalism. The result that an $AdS$/CFT correspondence is achieved, in this specific case, is due to the fact that the specific conformal transformation coupled with a change of coordinates maps the Schwarzschild solution to an extremal Segre non-null electromagnetic field solution, which is known to exhibit $AdS_2\times S^2$ topology in the near horizon. This essentially maps the question of the Schwarzschild/CFT correspondence to the Kerr/CFT correspondence applied to an extremal Reissner-N\"ordstom (RN) black hole \cite{Compere:2012jk}. However, the map chosen is conformal and we know from the above mentioned $c$-theorem \cite{ctzjet} that gravity in 2 dimensions should run to a conformal field theory, under the renormalization group flow, with center promotional to the horizon area of the parent four dimensional black hole \cite{carlip}. Also, under conformal transformations of the metric the form of $c$ and $L_0$ are unaffected, see \cite{ry} and references therein for a full discussion. In other words starting with a two dimensional spacetime, $g_{ab}$, with an unknown correspondence we are at liberty to conformally transform (wit respect to the pertaining CFT) to a black hole $g_{ab}=e^{\gamma \phi}g^{AdS_2}_{ab}$, where $g^{AdS_2}_{ab}$ has a known correspondence with calculable central charge that would be proportional to the conformal factor, which in turn contains quantum information about $g_{ab}$. The difficulty with the above paradigm is summarized as follows:
\begin{itemize}
\item A general four dimensional black hole does not necessary have a two dimensional representation.
\item Computation of the center depends on renormalization procedures of either the quantum CFT or the quantum energy momentum tensor. 
\item The auxiliary fields in the resulting CFT require physical boundary conditions rendering them finite on either the black hole horizon, or at asymptotic infinity or both. 
\end{itemize}
Taking advantage of these facts to (liberally) conformally map the Schwarzschild spacetime to one with global $AdS_2\times S^2$ topology in order to implement our CFT construction to compute the full asymptotic symmetry group provides us with a conformal window into what a full or complete Schwarzschild CFT construction may look like and we believe a novel step in the specific direction. 

As mentioned earlier, the vanishing surface gravity makes regulating the quantum charges of \eqr{eq:vir} cumbersome, thus leading to difficulty in using \eqr{eq:cf} for extremal RN geometry and/or Kerr. this is where the thermal Cardy formula \eqref{eq:tcf} is employed in addition with the Frolov-Thorne vacuum temperatures $T_L$ and $T_R$. In the case for the famous Kerr/CFT correspondence in the near extremal limit:
\begin{align}\label{eq:nefttemp}
T_L=\frac{(GM)^2}{2\pi J}\ ~\text{and}~\ T_R=\frac{\sqrt{(GM)^4-(GJ)^2}}{2\pi J}.
\end{align}
And in the extremal limit we have $T_L=\frac{1}{2\pi}$ and $T_R=0$, where $J=GM^2$. Additionally, since we have the assumption that $c_L=c_R=c=12J$ in the extremal case, the near horizon extremal Kerr (NHEK) entropy area law is obtained from \eqref{eq:tcf}:
\begin{align}
\label{eq:tcf2}
S_{BH}=\frac{\pi^2}{3}(c_LT_L+c_RT_R)=\frac{\pi^2}{3}cT_L=2\pi J.
\end{align} 
The above allows for the identification of a near horizon extremal Kerr (NHEK) temperature, 
\begin{equation}
T=T_L+T_R=\frac{1}{2\pi},\label{eq:cfttemp}
\end{equation}
which is in contrast to the vanishing Hawking temperature in this limit.

Now, while the implementation of the thermal Cardy formula is motivated by the absence of a properly regularized lowest Virasoro eigen mode \eqr{eq:vir}, a finite mode may be inferred, not from first principle of the ASG, from \eqref{eq:tcf2} by the identification
\begin{align}\label{eq:fttdef}
\frac{\partial S_{CFT}}{\partial \Q_0}=\frac{\partial S_{BH}}{\partial \Q_0}\equiv\frac{1}{T}\Rightarrow \Q_0=\frac{\pi^2}{6}c T^2.
\end{align}
Additionally, the general temperature $T=\frac{1}{2\pi}$ of \eqref{eq:cfttemp} may be obtained from \eqref{eq:fttdef} if the ASG contains a proper $SL(2,\mathbb{R})$ subgroup. Given the relationship of \eqr{eq:fttdef} we conlcude that $T$ in general should be unitless (for $\hbar=1$), which coincides with the Hawking temperature scaled by a finite time regulator $1/\kappa$ yielding: 
$T=\frac{1}{2\pi}$. This result extends smoothly to extremality (similarly to the identification found in \cite{Carlip:2011ax,ChangYoung:2012kd}). Now, since the extremal result $c=12J$ is consistent with the general expression $c=\frac{3A}{2\pi G}$, we may 
recast  $c,~T_L,~T_R$ in terms of more general black hole variables:
\begin{align}\label{eq:nexctlr}
c=\frac{3A}{2\pi G},~T_{L}=\frac{4(GM)^2}{A}~\text{and}~T_{R}=\frac{4\sqrt{(GM)^4-(GJ)^2}}{A}
\end{align}
and substituting \eqref{eq:nexctlr} into \eqr{eq:tcf} yields the standard Bekenstein-Hawking area Law $S_{BH}=\frac{A}{4G}$. Also, assuming a smooth extension to non-extremality, we have:
\begin{align}
T=T_{L}+T_{R}=\frac{4(GM)^2}{A}+\frac{4\sqrt{(GM)^4-(GJ)^2}}{A}=\frac{1}{2\pi}.
\end{align}
Such a procedure would provide a more wholesome calculation of near-extremal black hole entropy. 

Motivated by the above discussion, the aim of this note is to construct a $CFT$ dual for the near horizon Weyl rescaled geometry and compute the full ASG with an $SL(2,\mathbb{R})$ subgroup (which is absent in \cite{Shajiee2016,Sadeghi:2015uak,Ropotenko:2015egm}) and obtain its  entropy via the \emph{statistical} Cardy formula \eqr{eq:cf}. We will accomplish this by performing a Robinson and Wilczek near horizon dimensional reduction and construct a CFT out of the resulting two dimensional Kaluza-Klein field continent. The resulting two dimensional black hole is pure $AdS_2$ (not a non-null EM field) and we compute its full asymptotic symmetry group including lowest renormalized eigen mode. This should complete the near horizon quantum microstate study of the Weyl rescaled Schwarzschild geometry. Additionally, we discuss our results and premise within the conformal Weyl gravity formalism. As aforementioned, it is not entirely transparent why the use of a conformally rescaled geometry instead of the actual geometry is valid, since they are geometrodynamically different. However, the conformal Weyl gravity (CWG) paradigm (see \cite{Wheeler:2013ora} and the references therein) may straighten this implementation, since any two black holes related by a conformal transformation would belong to the same equivalence class of solutions to the respective CWG field equations. To engage in this topic, we solve the CWG vacuum field equations for a solution with $AdS_2\times S^2$ topology. We then study the resulting quantum microstates within an $AdS_2$/$CFT_1$ correspondence and compute its full ASG. Finally we discuss a dimensional reduction of CWG and see what possible avenues there are for extracting the central extension directly from its action principle. 
\section{On the Weyl Rescaled Schwarzschild CFT}\label{sec:SSCFT}
In this section we complete the study of the near horizon thermodynamics of the Weyl Rescaled Schwarzschild spacetime within a full computation of its near horizon ASG and quantum theory. In addition we demonstrate how both entropy and temperature are contained within one single respective formalism. 
\subsection{Geometry}\label{sec:SSCFTGeo}
We begin with the standard Schwarzschild spacetime metric:
\begin{align}
\label{eq:sss1.1}
\begin{split}
ds^2_{ss}=&-\left(1-\frac{2 G M}{r}\right)dt^2+\left(1-\frac{2 G M}{r}\right)^{-1}dr^2+r^2d\theta^2+r^2\sin^2\theta d\phi^2,
\end{split}
\end{align}
which can be transformed to exhibit $AdS_2\times S^2$ topology via the Weyl rescaling
\begin{align}
\label{eq:sstran1}
\begin{split}
g_{\mu\nu}\to& \Omega^{-2}g_{\mu\nu}\\
\text{for}~\Omega^{-2}=&\frac{(2GM)^2}{r^2}
\end{split}
\end{align}
followed by the map $r\to1/r$ and coordinate transformation:
\begin{align}
\label{eq:nhekcoor1}
r\to\lambda r,~t\to\frac{t}{(2GM)^2\lambda},
\end{align}
where the $(2GM)^2$ factor above is chosen to ensure the resulting black hole's horizon area is equivalent to the that of the original Schwarzschild spacetime in \eqr{eq:sss1.1}. Next, taking the limit $\lambda\to0$ we obtain the Weyl rescaled Schwarzschild spacetime: 
\begin{align}
\label{eq:sswrs}
ds^2=-\frac{r^2}{\dr^2}dt^2+\frac{\dr^2}{r^2}dr^2+{\dr^2}d\theta^2+{\dr^2}\sin^2\theta d\phi^2,
\end{align}
where $\dr^2=(2GM)^2$. The above Weyl rescaled Schwarzschild spacetime (WRSS) represents an extremal solution of \emph{Petrov Type D} and \emph{Segre Type} $\{(11),(1,1)\}$ (non-null)\footnote{We employ the standard spacetime classification notation of \cite{Stephani:2003tm}.} with global $AdS_2\times S^2$ topology, however we will tune it to near-extremality via the finite temperature/mass gauge:
\begin{align}
\label{eq:sswrs2}
ds^2_{WRSS}=-\frac{r^2-2GMr}{\dr^2}dt^2+\frac{\dr^2}{r^2-2GMr}dr^2+{\dr^2}d\theta^2+{\dr^2}\sin^2\theta d\phi^2,
\end{align}
where we have endowed the spacetime with an $ADM$ mass parameter and horizon $r_+=2GM$, which represents a finite excitation above extremality.  The above two line elements \eqref{eq:sswrs} and  \eqref{eq:sswrs2} are classically diffeomerophic \cite{Amsel:2009ev,Compere:2012jk}, as they both exhibit identical canonical Riemann tensors and covariant derivatives of their canonical Riemann tensors to tenth order according to the Cartan-Karlhede spacetime equivalence algorithm/theorem \cite{Stephani:2003tm,Karlhede1980,Karlhede2006}:
\begin{align}
R\indices{^{\eqref{eq:sswrs}}_{\mu\alpha\nu\beta}}=&R\indices{^{\eqref{eq:sswrs2}}_{\mu\alpha\nu\beta}}\\
\nabla_\rho R\indices{^{\eqref{eq:sswrs}}_{\mu\alpha\nu\beta}}=&\nabla_\rho R\indices{^{\eqref{eq:sswrs2}}_{\mu\alpha\nu\beta}}=0.
\end{align}
Since the first covariant derivative of both curvature tensors is identically zero, we need not list additional. However they induce different quantum theories in their respective near horizons and the excitation above extremality will aid substantially in our $CFT$ construction. 
\subsection{Quantum Fields in WRSS Spacetime}\label{sec:qftsswrs}
Similarly to the background discussed in the introduction, we will now perform a semi-classical analysis in the near horizon of \eqref{eq:sswrs2} in order to extract the two dimensional field content relevant for our $CFT$ construction. We do this by probing the spacetime with a minimally coupled scalar, for which we know the quantum effective action:
\begin{align}
\label{eq:seffa}
S_{eff}\sim\frac{\bar\beta^\psi}{16\pi}\int d^2x\sqrt{-g^{(2)}}R^{(2)}\frac{1}{\square_{g^{(2)}}}R^{(2)}+\cdots,
\end{align}
where  $\bar\beta^\psi=\frac{const}{G}$ is the Weyl anomaly coefficient \cite{tseytlin89,polyak}. Our goal will be to determine the respective effective theory and value of $\bar\beta^\psi$ within an $s$-wave approximation. This is reasonable if we assume our scalar probe has gravitational origin, i.e. \eqref{eq:sswrs2} exhibits a Kaluza-Klein decomposition:
\begin{align}
\label{eq:wrsskk}
\begin{split}
ds^2=&-f(r)dt^2+f(r)^{-1}dr^2+\alpha^2 e^{-2\psi(r)}\left[d\theta^2+\sin^2\theta \left(d\phi+Adt\right)^2\right]\\
=&ds^2_{2D}+\alpha^2 e^{-2\psi(r)}\left[d\theta^2+\sin^2\theta \left(d\phi+A_\mu dx^\mu\right)^2\right]
\end{split}
\end{align}
in terms of a two dimensional black hole coupled to a two dimensional real scalar and two dimensional $U(1)$ gauge field. In the case of \eqref{eq:sswrs2}, the only allowable two dimensional gauge field couplings are given by linear phase shifts of the form $\phi\to\phi+At$, and thus are trivial. Even for spacetimes that do not have global spherical symmetry, the $s$-wave approximation is still valid, since all of the gravitational dynamics seem to be contained in this regime \cite{strom1}\footnote{In \cite{Button:2010kg} it was shown that $lm$ terms above $00$ decay exponentially fast in time by analyzing the asymptotic behavior of its field equation for an axisymmetric spacetime.}.

To begin extracting the relevant two dimensional near horizon field content, we consider a four dimensional massless free scalar field in the background of \eqref{eq:sswrs2}:
\begin{align}
\label{eq:freescalar4}
\begin{split}
S_{free}=&\frac12\int d^4x\sqrt{-g}g^{\mu\nu}\partial_{\mu}\varphi\partial_\nu\varphi\\
=&-\frac12\int d^4x\,\varphi\left[\partial_\mu\left(\sqrt{-g}g^{\mu\nu}\partial_\nu\right)\right]\varphi\\
=&-\frac12\int d^4x\,\varphi\left[\partial_t\left( -\dr^2\sin{\theta}\frac{\dr^2}{r^2-2GMr}\partial_t\right)+\partial_r\left(\dr^2\sin{\theta}\frac{r^2-2rGM}{\dr^2}\partial_r\right)\right.\\
&\left.+\partial_\theta\left( \dr^2\sin{\theta}\frac{1}{\dr^2}\partial_\theta\right)+\partial_\phi\left( \dr^2\sin{\theta}\frac{1}{\dr^2\sin^2{\theta}}\partial_\phi\right)\right]\varphi.
\end{split}
\end{align}
The above functional can be reduced to a two dimensional theory by expanding $\varphi$ in terms of spherical harmonics via:
\begin{align}
\label{eq:sphdecom}
\varphi(t,r,\theta,\phi)=\sum_{lm}\varphi_{lm}(r,t)Y\indices{_l^m}(\theta,\phi),
\end{align}
where $\varphi_{lm}$ is an interacting complex two dimensional scalar field. Next, integrating out angular degrees of freedom, performing a change of coordinates to tortoise coordinates $dr^*=f(r)dr$ and examining the region $r\sim r_+$, reduces the two dimensional action even more. This is due to the fact that interacting, mixing and potential terms ($\sim l(l+1)\ldots$) are weighted by a factor of $f(r(r*))\sim e^{2\kappa r^{*}}$, which vanishes exponentially fast in the region near the horizon $r\sim r_+$. This leaves us with a collection of infinite massless scalar fields with action functional:
\begin{align}
\label{eq:nhwrss}
\begin{split}
S=&-\frac{\dr^2}{2}\int d^2x\;\varphi^{*}_{lm}\left[-\frac{1}{f(r)}\left(\partial_t\right)^2+\partial_rf(r)\partial_r\right]\varphi_{lm}\\
=&-\frac{\ell^2}{2}\int d^2x\;\varphi^{*}_{lm}D_{\mu}\left[\sqrt{-g^{(2)}}g^{\mu\nu}_{(2)}D_{\nu}\right]\varphi_{lm},
\end{split}
\end{align}
where $D_\mu=\partial_\mu-im\A_\mu$ is the gauge covariant derivative. Thus, we arrive at the Robinson and Wilczek two dimensional analog (RW2DA) fields for the WRSS spacetime solution given by:
\begin{align}
\label{eq:2drwamet}
\g{^{(2)}_\mu_\nu}=&\left(\begin{array}{cc}-f(r) & 0 \\0 & \frac{1}{f(r)}\end{array}\right)&f(r)=\frac{r^2-2GMr}{\dr^2}
\end{align}
with trivial $U(1)$ gauge field.
\begin{align}
\label{eq:rw2dgf}
\A=\A_tdt=constant=0.
\end{align}
Again, given the WRSS ansatz \eqr{eq:sswrs2}, it is not surprising that the only relevant physical fields in the region $r\sim r_+$ are the above RW2DAs, with trivial gauge sector. This is due to its spherical symmetry and we will continue in the next section with a holographic semiclassical analysis of $\g{^{(2)}_\mu_\nu}$, $\A$ and $\varphi_{lm}$ to obtain the dual quantum CFT in the near horizon regime.

The quantum effective functional of \eqr{eq:nhwrss}, to $s$-wave $\varphi_{00}=\sqrt{\frac{6}{G}}\psi$ for unitless\footnote{The factor $\sqrt{6}$ is chosen to coincide with the normalization of \eqr{eq:seffa} for the gravitational sector of the effective action.} $\psi$, is obtained via path integrating over $\psi$, which amounts to a zeta-function regularization of the functional determinant of $D_{\mu}\left[\sqrt{-g^{(2)}}g^{\mu\nu}_{(2)}D_{\nu}\right]$. In general it is comprised of the two parts \cite{Leutwyler:1984nd,isowill}:
\begin{align}\label{eq:nhpcft}
\Gamma=&\Gamma_{grav}+\Gamma_{U(1)},
\end{align}
where
\begin{align}\label{eq:nhnloc1}
\begin{split}
\Gamma_{grav}=&\frac{\dr^2}{16\pi G}\int d^2x\sqrt{-g^{(2)}}R^{(2)}\frac{1}{\square_{g^{(2)}}}R^{(2)},\\
\Gamma_{U(1)}=&\frac{3 e^2 \dr^2}{\pi G}\int \mc{F}\frac{1}{\square_{g^{(2)}}}\mc{F}
\end{split}
\end{align}
and in concurrence with \eqr{eq:seffa} for $\bar\beta^\psi=\frac{\ell^2}{G}$. $R^{(2)}$ above is the Ricci scalar curvature obtained from $\g{^{(2)}_{\mu\nu}}$ and $\mc{F}=d\mc{A}$ is the $U(1)$ invariant curvature two form. Naturally, $\Gamma_{U(1)}=0$ in \eqref{eq:nhpcft} for the case of WRSS, but we will keep it in our analysis for completeness and use in later sections. Next, we wish to restore locality in \eqref{eq:nhnloc1}, which may be done by introducing auxiliary scalars $\Phi$ and $B$ such that:
\begin{align}\label{eq:afeqm}
\square_{g^{(2)}} \Phi=R^{(2)}~\mbox{and}~\square_{g^{(2)}} B=\epsilon^{\mu\nu}\partial_\mu \A_\nu,
\end{align}
which for general $f(r)$ and $\A_t(r)$ have the form:
\begin{align}
\label{eq:afsol1.1}
\bs
-\frac{1}{f(r)}\partial_t^2\Phi+\partial_rf(r)\partial_r\Phi=&R^{(2)}\\
-\frac{1}{f(r)}\partial_t^2B+\partial_rf(r)\partial_rB=&-\partial_r A_t,
\es
\end{align}
with general solutions:
\begin{align}
\label{eq:afsol}
\bs
\Phi(t,r)=&\alpha_1 t+\int dr\frac{\alpha_2-f'(r)}{f(r)}\\
B(t,r)=&\beta_1 t+\int dr\frac{\beta_2-\A_t(r)}{f(r)},
\es
\end{align}
where $\alpha_i$ and $\beta_i$ are integration constants. Using \eqref{eq:afeqm} in \eqref{eq:nhnloc1} yields a near horizon Liouville type $CFT$ of the form:
\begin{align}\label{eq:nhlcft}
\begin{split}
S_{NHCFT}=&\frac{\dr^2}{16\pi G}\int d^2x\sqrt{-g^{(2)}}\left\{-\Phi\square_{g^{(2)}}\Phi+2\Phi R^{(2)}\right\}\\
&+\frac{3 e^2 \dr^2}{\pi G}\int d^2x\sqrt{-g^{(2)}}\left\{-B\square_{g^{(2)}}B+2B \left(\frac{\epsilon^{\mu\nu}}{\sqrt{-g^{(2)}}}\right)\partial_\mu A_\nu\right\}
\end{split}
\end{align}
\subsection{Asymptotic Symmetries}\label{sec:wrssas}
Now, we will compute the non trivial asymptotic symmetries of relevance for the gravitational sector of \eqr{eq:2drwamet} with large $r$ behavior defined by
\begin{align}\label{eq:wrss2dwa}
\g{^{(0)}_\mu_\nu}=&
\left(
\begin{array}{cc}
-\frac{r^2}{\dr^2}+\frac{2 r G M}{\dr^2}+\mathcal{O}\left(\left(\frac{1}{r}\right)^3\right)& 0 \\
 0 & \frac{\dr^2}{r^2}+\mathcal{O}\left(\left(\frac{1}{r}\right)^3\right) 
\end{array}
\right)
\end{align}
which yield an asymptotically $AdS_2$ configuration with Ricci Scalar, $R=-\frac{2}{l^2}+O\left(\left(\frac{1}{r}\right)^1\right)$. We couple this with the following metric fall-off conditions:
\begin{align}\label{eq:mbc}
\delta g_{\mu\nu}=
\left(
\begin{array}{cc}
    \mathcal{O}\left(\left(\frac{1}{r}\right)^3\right)&
   \mathcal{O}\left(\left(\frac{1}{r}\right)^0\right) \\
 \mathcal{O}\left(\left(\frac{1}{r}\right)^0\right) &
\mathcal{O}\left(r\right) 
\end{array}
\right),
\end{align}
which are preserved by the following set of asymptotic diffeomorphisms:
\begin{align}\label{eq:dpr}
\chi=-C_1\frac{r \xi(t)}{r-2GM}\partial_t+C_2r\xi'(t)\partial_r,
\end{align}
where $\xi(t)$ is an arbitrary function and $C_i$ are arbitrary normalization constants. Switching to conformal light cone coordinates given by:
\begin{align}
x^\pm=t\pm r^*
\end{align}
and transforming \eqref{eq:dpr}, we obtain:
\begin{align}
\label{eq:lcdiff}
\chi^\pm=\frac{\left(-C_1r(r^*) \xi(x^+,x^-)\pm C_2\ell^2 \xi'(x^+,x^-)\right)}{r(r^*)-2 G M},
\end{align}
which are smooth on the asymptotic boundary. 
\subsection{Energy-Momentum and Central Charge}\label{sec:thermo}
The energy-momentum tensor is defined in the usual way:
\begin{align}
\label{eq:emt}
\begin{split}
\left\langle T_{\mu\nu}\right\rangle=&\frac{2}{\sqrt{-g^{(2)}}}\frac{\delta S_{NHCFT}}{\delta g\indices{^{(2)}^\mu^\nu}}\\
=&\frac{\dr^2}{8\pi G}\left\{\partial_\mu\Phi\partial_\nu\Phi-2\nabla_\mu\partial_\nu\Phi+g\indices{^{(2)}_\mu_\nu}\left[2R^{(2)}-\frac12\nabla_\alpha\Phi\nabla^\alpha\Phi\right]\right\}
\end{split}
\end{align}
Next, we substitute the general solution \eqr{eq:afsol} into \eqr{eq:emt} and adopt modified Unruh Vacuum boundary conditions (MUBC) \cite{unruh}
\begin{align}
\label{eq:ubc}
\begin{cases}
\left\langle T_{++}\right\rangle=0&r\rightarrow\infty,~\ell\rightarrow\infty\\
\left\langle T_{--}\right\rangle=0&r\rightarrow r_+
\end{cases},
\end{align}
which infer the general behavior:
\begin{align}
\begin{cases}
f(r)=0&r\to r_+\\
f(r)=0&\ell\to \infty\\
\end{cases}.
\end{align}
These results allows the determination of the integration constants $\alpha_i$, yielding:
\begin{align}
\bs
\alpha_1=&-\alpha_2=\frac12f'(r_+)
\es
\end{align}
and thus, determining the EMT. The above EMT exhibits a Weyl (trace) anomaly given by:
\begin{align}
\label{eq:tra}
\left\langle T\indices{_\mu^\mu}\right\rangle=-\frac{\bar\beta^\psi}{4\pi}R^{(2)},
\end{align}
which uniquely determines the value of central charge via \cite{cft}:
\begin{align}
\label{eq:center1}
\frac{c}{24\pi}=\frac{\bar\beta^\psi}{4\pi}\Rightarrow c=6\ell^2/G=\frac{3A}{2\pi G},
\end{align}
in agreement with \eqref{eq:nexctlrintro}. Additionally, due to the use of the MUBC, the asymptotic boundary of interest (and to $\mathcal{O}(\frac{1}{\ell})^2$, which we denote by the single limit $x^+\to\infty$) the EMT is dominated by one holomorphic component $\left\langle T_{--}\right\rangle$. Expanding this component in terms of the boundary fields \eqr{eq:wrss2dwa} and computing its response to the asymptotic symmetry yields:
\begin{align}
\delta_{\chi^-}\left\langle T_{--}\right\rangle=\chi^-\left\langle T_{--}\right\rangle'+2\left\langle T_{--}\right\rangle\left(\chi^-\right)'+\frac{c}{24\pi}\left(\chi^-\right)'''+\mathcal{O}\left(\left(\frac{1}{r}\right)^3\right),
\end{align}
where prime denotes temporal derivatives. The above implies that $\left\langle T_{--}\right\rangle$ transforms asymptotically as the EMT of a one dimensional $CFT$ with center \eqref{eq:center1}.
\subsection{Full Asymptotic Symmetry Group}\label{sec:vafullasg}
Next, we compute the full ASG or charge algebra by compactifying the $x^-$ coordinate to a circle from $0\to4\pi\ell^2/r_+$ and defining the asymptotic conserved charge:
\begin{align}
\label{eq:asycharge}
\mathcal{Q}_n=\lim_{x^+\to\infty}\int dx^\mu\left\langle T_{\mu\nu}\right\rangle\chi^\nu_n,
\end{align}
where we replaced $\xi(x^+,x^-)$ with circle diffieomorphisms $\frac{e^{-in(r_+/2\ell^2)x^\pm}}{r_+/2\ell^2}$ in \eqr{eq:lcdiff} and the $C_i$'s are fixed by requiring the $\chi^-_n$ to form an asymptotic centerless Witt or $Diff(S^1)$ subalgebra:
\begin{align}
i\left\{\chi^-_m,\chi^-_n\right\}=(m-n)\chi^-_{m+n}.
\end{align}
Now, calculating the canonical response of $\mathcal{Q}_n$ with respect to the asymptotic symmetry yields:
\begin{align}
\label{eq:ca}
\delta_{\chi^-_m}\Q_n=\left[\Q_m,\Q_n\right]=(m-n)\Q_n+\frac{c}{12}m\left(m^2-1\right)\delta_{m+n,0},
\end{align}
i.e. the asymptotic quantum generators form a centrally extended Virasoro algebra with central chrage \eqr{eq:center1} and computable non-zero lowest eigen-mode: 
\begin{align}
\label{eq:q0}
\Q_0=\frac{\dr^2}{4 G}=\frac{A}{16\pi G}.
\end{align}
\subsection{$AdS_2/CFT_1$ and WRSS Thermodynamics}\label{sec:wrssent}
Summarizing from the previous section, we showed that the WRSS is holographically dual to a CFT with center:
\begin{align}\label{eq:cre}
c&=\frac{3A}{2\pi G}
\end{align}
and lowest Virasoro eigenmode 
\begin{align}
\Q_0&=\frac{A}{16\pi G}.
\end{align}
Using this results in the statistical (not thermal) Cardy Formula \eqr{eq:cf},
\begin{align}
S=2\pi\sqrt{\frac{c\Q_0}{6}}=\frac{A}{4G}=\frac{\dr^2}{4G},
\end{align}
which is in agreement with the Bekenstein-Hawking area law \eqr{eq:htaen} for the Schwarzschild spacetime \eqr{eq:sss1.1}.

Next, to compute the WRSS black hole temperature we turn our attention back to the EMT \eqr{eq:emt}, 
which on the horizon $r\to r_+$ is dominated by one holomorphic component:
\begin{align}
\label{eq:hhf}
\left\langle T_{++}\right\rangle=-\frac{\dr^2}{32 \pi G }f'\left(r^+\right)^2
\end{align}
which is precisely the Hawking flux ($HF$, radiation flux $\sim T\indices{_r^t}$) of the WRSS metric, weighted by the central charge \eqr{eq:center1}:
\begin{align}
\label{eq:htfhf}
\left\langle T_{++}\right\rangle=cHF=-c\frac{\pi}{12}\left(T_H\right)^2
\end{align}
with Hawking temperature\cite{Jinwu,caldarelli:1999xj}
\begin{align}
T_H=\frac{f'\left(r^+\right)}{4 \pi }.
\end{align}
The above are interesting results, as it demonstrates that the $AdS_2/CFT_1$ correspondence constructed here contains information about both black hole entropy and black hole temperature. Though, we should note that the $\left\langle T_{++}\right\rangle$ component in the respective limit is not precisely the Hawking flux of the four dimensional parent black hole, yet having prior knowledge of the central extension, it is possible to read off the relevant information from the correspondence.
\section{Canonical Microstates From the ASG and Conformal Weyl Gravity}\label{sec:CWGBH}
As mentioned in the introduction we turn our attention now to the computation of the full ASG of a back hole originating from the Conformal Weyl gravity (CWG) paradigm. The CWG action function is given by:
\begin{align}\label{eq:CWGA}
S_{CWG}=\alpha_{g}\int d^4x\sqrt{-g}W^{\alpha\mu\beta\nu}W_{\alpha\mu\beta\nu},
\end{align}
where $\alpha_{g}$ is a unit-less coupling and $W_{\alpha\mu\beta\nu}$ is the Weyl tensor defined in its usual way. The above action is invariant under conformal transformations and thus \eqr{eq:sss1.1} and \eqr{eq:sswrs2} belong to the same equivalence class of solutions within this formalism. The unit-less coupling makes CWG a naively attractive candidate for pursuing quantum gravity, it also includes cosmological dynamics as part of its vacuum, however the fourth order nature and unitarity issues endow it with some difficulties. It is nonetheless an attractive theory and includes all the standard solar-system tests of general relatively and warrants investigation on its own right. In this section will construct a near horizon CFT dual to a black hole belonging uniquely to the CWG vacuum solution space. We compute the resulting implied ASG and thermodynamics. This analysis strengthens the premise of equivalent black hole thermodynamic systems related by a conformal transformation of the parent black holes. 
\subsection{CWG Geometry}\label{sec:CWGsol}
The vacuum equation of motion of CWG resulting from varying \eqr{eq:CWGA} with respect to the inverse metric reads:
\begin{align}\label{eq:cwgeqm}
2\nabla_\alpha\nabla_\beta W\indices{_\mu^\alpha^\beta_\nu}+R_{\alpha\beta}W\indices{_\mu^\alpha^\beta_\nu}=0
\end{align}
We will solve the above equation starting with the initial Kaluza-Klein ansatz:
\begin{align}
ds^2=&K_1(\theta)\left[-f(r)dt^2+f(r)^{-1}dr^2+\dr^2d\theta^2\right]+\frac{\dr^2sin^2{\theta}}{K_2(\theta)} \left(d\phi-A_\mu(r)dx^\mu\right)^2,
\end{align}
which includes a particular interesting solution (CWGS) with finite mass/temperature: 
\begin{align}
f(r)=&\frac{r^2-2GMr}{\ell^2}&A_\mu(r)dx^\mu=-\frac{r-2GM}{\dr^2}dt~\text{and}\\
K_1(\theta)=&\sin{\theta}&K_2(\theta)=\sin{\theta}
\end{align}
and line element:
\begin{align}\label{eq:mysol}
ds^2_{CWGS}=&\sin{\theta}\left[-\frac{r^2-2GMr}{\ell^2}dt^2+\frac{\ell^2}{r^2-2GMr}dr^2+\dr^2d\theta^2\right]+\dr^2sin{\theta} \left(d\phi+\frac{r-2GM}{\dr^2}dt\right)^2.
\end{align}
The above line element exhibits local $AdS_2\times S^2$ topology and non trivial two dimensional $U(1)$ gauge potential. It is a black hole of global \emph{Petrov Type O} and \emph{Segre Type} $\{(111),1\}$ (perfect fluid), with Kretschmann Invariant: 
\begin{align}
R_{\mu\nu\alpha\beta}R^{\mu\nu\alpha\beta}=\frac{15 \csc^6{\theta}}{4\dr^2}
\end{align}
and horizon located at $r_+=2GM$ with area $A=4\pi\dr^2$. We computed and listed the classification here for completeness, but are not in general interested in the novelty of \eqr{eq:mysol}, but only in the near-horizion quantum theory. The $AdS_2$ factor and finite mass/temperature in \eqr{eq:mysol} will aid (as before) considerably in our next analytical pursuits. 
\subsection{Quantum Fields in CWGS}\label{sec:qftcwgs}
Following from Section~\ref{sec:qftsswrs}, we will similarly begin by probing \eqr{eq:mysol} with a free scalar field:
\begin{align}
\label{eq:freescalarCWGS}
\begin{split}
S_{free}=&\frac12\int d^4x\sqrt{-g}g^{\mu\nu}\partial_{\mu}\varphi\partial_\nu\varphi\\
=&-\frac12\int d^4x\,\varphi\left[\partial_\mu\left(\sqrt{-g}g^{\mu\nu}\partial_\nu\right)\right]\varphi\\
=&-\frac12\int d^4x\,\varphi\left[\partial_t\left( -\dr^2\sin{\theta}\frac{\dr^2}{r^2-2GMr}\partial_t\right)+\partial_r\left(\dr^2\sin{\theta}\frac{r^2-2rGM}{\dr^2}\partial_r\right)\right.\\
&\left.+\partial_\theta\left( \dr^2\sin{\theta}\frac{1}{\dr^2}\partial_\theta\right)+\partial_\phi\left(\left\{ \sin{\theta}-\dr^2\sin{\theta}\frac{r^2}{\dr^4}\frac{\dr^2}{r^2-2GMr}\right\}\partial_\phi\right)\right.\\
&\left.+2\partial_t\left(-\dr^2\sin{\theta}\frac{r}{\dr^2}\frac{\dr^2}{r^2-2GMr}\partial_\phi\right)\right]\varphi.
\end{split}
\end{align}
Again, we reduce to a two dimensional theory by expanding $\varphi$ in terms of spherical harmonics \eqr{eq:sphdecom}, integrating out angular degrees of freedom, changing to tortoise coordinates and taking the limit $r\sim r_+$, which leaves us with a (similar to \eqr{eq:nhwrss}) collection of infinite massless scalar fields with action functional:
\begin{align}
\begin{split}
S=&-\frac{\dr^2}{2}\int d^2x\;\varphi^{*}_{lm}\left[-\frac{1}{f(r)}\left(\partial_t-imA_t(r)\right)^2+\partial_rf(r)\partial_r\right]\varphi_{lm}\\
=&-\frac{\ell^2}{2}\int d^2x\;\varphi^{*}_{lm}D_{\mu}\left[\sqrt{-g^{(2)}}g^{\mu\nu}_{(2)}D_{\nu}\right]\varphi_{lm},
\end{split}
\end{align}
Thus, we arrive at the RW2DA fields for the CWGS spacetime given by:
\begin{align}
\label{eq:2drwacwg}
\g{^{(2)}_\mu_\nu}=&\left(\begin{array}{cc}-f(r) & 0 \\0 & \frac{1}{f(r)}\end{array}\right)&f(r)=\frac{r^2-2GMr}{\dr^2}
\end{align}
and $U(1)$ gauge field.
\begin{align}
\label{eq:rw2dgf1.1}
\A=A_t(r)dt=-\frac{r-2GM}{\dr^2}dt.
\end{align}
At this point we follow the same steps as in the previous section going from \eqr{eq:nhpcft} to \eqr{eq:nhlcft} and yielding again a Liouville type near horizon $CFT$ for the RW2DA of the CWGS spacetime:
\begin{align}
\begin{split}
S_{NHCFT}=&\frac{\dr^2}{16\pi G}\int d^2x\sqrt{-g^{(2)}}\left\{-\Phi\square_{g^{(2)}}\Phi+2\Phi R^{(2)}\right\}\\
&+\frac{3 e^2 \dr^2}{\pi G}\int d^2x\sqrt{-g^{(2)}}\left\{-B\square_{g^{(2)}}B+2B \left(\frac{\epsilon^{\mu\nu}}{\sqrt{-g^{(2)}}}\right)\partial_\mu A_\nu\right\}
\end{split}
\end{align}
\subsection{CWGS Asymptotic Symmetries}
We will now focus on the non trivial asymptotic symmetries of relevance for the RW2DA fields \eqr{eq:2drwacwg} and \eqr{eq:rw2dgf1.1} with large $r$ behavior defined by:
\begin{align}\label{eq:cwgs2dwa1}
\g{^{(0)}_\mu_\nu}=&
\left(
\begin{array}{cc}
-\frac{r^2}{\dr^2}+\frac{2 r G M}{\dr^2}+\mathcal{O}\left(\left(\frac{1}{r}\right)^3\right)& 0 \\
 0 & \frac{\dr^2}{r^2}+\mathcal{O}\left(\left(\frac{1}{r}\right)^3\right) 
\end{array}
\right)\\
\mathcal{A}\indices{^{(0)}_t}=&\frac{r}{\ell^2}+\mathcal{O}\left(\frac{1}{r}\right)^3
\end{align}
which yield an asymptotically $AdS_2$ configuration with Ricci Scalar, $R=-\frac{2}{l^2}+O\left(\left(\frac{1}{r}\right)^1\right)$. We couple this with the following metric and gauge field fall-off conditions:
\begin{align}
\delta g_{\mu\nu}=&
\left(
\begin{array}{cc}
    \mathcal{O}\left(\left(\frac{1}{r}\right)^3\right)&
   \mathcal{O}\left(\left(\frac{1}{r}\right)^0\right) \\
 \mathcal{O}\left(\left(\frac{1}{r}\right)^0\right) &
\mathcal{O}\left(r\right) 
\end{array}
\right)\\
\delta \mathcal{A}=&\mathcal{O}\left(\frac{1}{r}\right)^0
\end{align}
which are again preserved by the following set of asymptotic diffeomorphisms:
\begin{align}\label{eq:dpr1.1}
\chi=-C_1\frac{r \xi(t)}{r-2GM}\partial_t+C_2r\xi'(t)\partial_r,
\end{align}
The variation of the gauge field with respect to the above diffeomorphisms yields $\delta_\chi \mathcal{A}=\mathcal{O}\left(\frac{1}{r}\right)^0$ and thus, $\delta_\chi$ is trivially elevated to a total symmetry 
\begin{align}
\label{eq:totsym}
\delta_{\chi}\to\delta_{\chi+\Lambda}
\end{align}
of the action. Finally, again switching to conformal light cone coordinates and transforming \eqref{eq:dpr1.1}, we obtain:
\begin{align}
\chi^\pm=\frac{\left(-C_1r(r^*) \xi(x^+,x^-)\pm C_2\ell^2 \xi'(x^+,x^-)\right)}{r(r^*)-2 G M},
\end{align}
which are still smooth on the relevant asymptotic boundary. 
\subsection{Energy-Momentum and CWGS Central Charge}
The energy-momentum is given by \eqref{eq:emt} and the $U(1)$ current is defined as:
\begin{align}
\label{eq:emt1.1}
\begin{split}
\left\langle J^{\mu}\right\rangle=&\frac{1}{\sqrt{-g^{(2)}}}\frac{\delta S_{NHCFT}}{\delta A_\mu}=\frac{6 e^2 \ell^2}{\pi}\frac{1}{\sqrt{-g^{(2)}}}\epsilon^{\mu\nu}\partial_\nu B.
\end{split}
\end{align}
Again, substituting the general solution \eqr{eq:afsol} into \eqr{eq:emt1.1} and employing MUBC on both currents above:
\begin{align}
\label{eq:ubc1.1}
\begin{cases}
\left\langle T_{++}\right\rangle=\left\langle J_{+}\right\rangle=0&r\rightarrow\infty,~\ell\rightarrow\infty\\
\left\langle T_{--}\right\rangle=\left\langle J_{-}\right\rangle=0&r\rightarrow \epsilon
\end{cases}
\end{align}
we obtain the general behavior:
\begin{align}
\begin{cases}
f(r)=0&r\to \epsilon\\
f(r)=A_t(r)=0&\ell\to \infty\\
\end{cases}.
\end{align}
These results allow the determination of the integration constants $\alpha_i$ and $\beta_i$:
\begin{align}
\bs
\alpha_1=&-\alpha_2=\frac12f'(\epsilon) \\
\beta_1=&-\beta_2=\frac12\A_t(\epsilon) 
\es
\end{align}
and thus, determining the EMT and $U(1)$ current. As before, computing the Weyl (trace) anomaly:
\begin{align}
\label{eq:tra1.1}
\left\langle T\indices{_\mu^\mu}\right\rangle=-\frac{\bar\beta^\psi}{4\pi}R^{(2)},
\end{align}
determines uniquely the value of central charge:
\begin{align}
\label{eq:center11.1}
\frac{c}{24\pi}=\frac{\bar\beta^\psi}{4\pi}\Rightarrow c=6\ell^2/G=\frac{3A}{2\pi G},
\end{align}
which again agrees with \eqref{eq:nexctlrintro}. Additionally and due to the use of the MUBC, the EMT is dominated by one holomorphic component $\left\langle T_{--}\right\rangle$ on the asymptotic boundary $x^+\to\infty$. Expanding this component in terms of the boundary fields \eqr{eq:cwgs2dwa1} and computing its response to a total symmetry yields:
\begin{align}
\begin{cases}
\delta_{\chi^-+\Lambda}\left\langle T_{--}\right\rangle=\chi^-\left\langle T_{--}\right\rangle'+2\left\langle T_{--}\right\rangle\left(\chi^-\right)'+\frac{c}{24\pi}\left(\chi^-\right)'''+\mathcal{O}\left(\left(\frac{1}{r}\right)^3\right)\\
\delta_{\chi^-+\Lambda}\left\langle J_{-}\right\rangle=\mathcal{O}\left(\left(\frac{1}{r}\right)^3\right)
\end{cases}
\end{align}
and thus, $\left\langle T_{--}\right\rangle$ transforms again asymptotically as the EMT of a one dimensional $CFT$ with center \eqref{eq:center11.1}.
\subsection{Full ASG for CWGS}
Having done most of the leg work now, this section parallels Section~\ref{sec:vafullasg} very closely. To compute the charge algebra we compactify the $x^-$ coordinate to a circle from $0\to4\pi\ell^2/r_+$ and define the asymptotic conserved charge:
\begin{align}
\label{eq:asychargewrsscg1}
\mathcal{Q}_n=\lim_{x^+\to\infty}\int dx^\mu\left\langle T_{\mu\nu}\right\rangle\chi^\nu_n,
\end{align}
Next, making the same substations and requirements as in Section~\ref{sec:vafullasg} we obtain:
\begin{align}
\label{eq:ca1.1}
\delta_{\chi^-_m}\Q_n=\left[\Q_m,\Q_n\right]=(m-n)\Q_n+\frac{c}{12}m\left(m^2-1\right)\delta_{m+n,0},
\end{align}
i.e. the asymptotic quantum generators of the CWGS spacetime form a centrally extended Virasoro algebra with central chrage \eqr{eq:center11.1} and computable non-zero lowest eigen-mode: 
\begin{align}
\label{eq:q01.1}
\Q_0=\frac{\dr^2}{4 G}=\frac{A}{16\pi G}.
\end{align}
\subsection{$AdS_2/CFT_1$ and CWGS Thermodynamics}
Substituting \eqr{eq:q01.1} and \eqr{eq:center11.1} into the statistical Cardy Formula \eqr{eq:cf}, we again reproduce the Bekenstein-Hawking area law for the CWGS:
\begin{align}
S=2\pi\sqrt{\frac{c\Q_0}{6}}=\frac{A}{4G}=\frac{\dr^2}{4G},
\end{align}
Next, to compute the CWGS black hole temperature we turn our attention back to \eqref{eq:emt}, since we need only consider the gravitational sector, 
which on the horizon $r\to r_+$ is dominated by one holomorphic component:
\begin{align}
\label{eq:hhf1.1}
\left\langle T_{++}\right\rangle=-\frac{\dr^2}{32 \pi G }f'\left(r^+\right)^2
\end{align}
from which we again are able to read off the Hawking flux, given prior knowledge of the central charge \eqr{eq:center11.1}:
\begin{align}
\label{eq:htfhf1.1}
\left\langle T_{++}\right\rangle=cHF=-c\frac{\pi}{12}\left(T_H\right)^2
\end{align}
and thus we obtain the Hawking temperature
\begin{align}
T_H=\frac{f'\left(r^+\right)}{4 \pi }.
\end{align}
Once more, the above analysis demonstrates that our constructed $AdS_2/CFT_1$ correspondence contains information about both black hole entropy and black hole temperature.
\section{Discussion and Concluding Remarks}
In this paper we have shown that the Weyl rescaled Schwarzschild spacetime \eqr{eq:sswrs2} and the conformal Weyl gravity solution \eqr{eq:mysol} are both holographically dual to a Liouville CFT \eqr{eq:nhlcft} and have also computed their full ASG in their respective near horizon regimes. These parallel results strengthens the use of conformal Weyl rescalings to study the thermodynamics of the Schwarzschild spacetime, since both belong to the same equivalence class of solutions within the CWG formalism. Additionally we have performed our analysis by analyzing specific black holes of the solution spaces of Einstein and Conformal Weyl gravities. Equally interesting would be to perform the ASG calculation at the action level of CWG. That is starting with a general spacetime of the form: 
\begin{align}
ds^2=-f(r)dt^2+\frac{1}{f(r)}dr^2+e^{-2\psi(r)}\dr^2d\theta^2+e^{-2\psi(r)}\dr^2\sin^2\theta d\phi^2,
\end{align}
substituting into \eqr{eq:CWGA} and integrating out angular degrees of freedom yielding a two dimensional Conformal Weyl dilation gravity:
\begin{align}
S_{CWDG}=\frac{16\pi\dr^2\alpha_g}{3}\int d^2\sqrt{-g}e^{-2\psi(r)}\left[-\frac{R}{2}+\square\psi-\frac{e^{2\psi}}{\dr^2}\right]^2,
\end{align}
up to total derivatives. The above two dimensional theory is not conformally invariant anymore (as compared to its parent CWG) and signals the presence of an anomaly. A holographic regularization of its boundary stress tensor dynamics would be interesting and should reveal the details of the respective anomaly and how it relates to the centers derived in the previous sections. This would be a good starting point for additional future interesting work on the central charge of CWG. 

Since we are on the topic of holographic regularization, our derived zero-mode in \eqr{eq:q01.1} computed from \eqr{eq:asychargewrsscg1} in terms of a holographic regulated energy momentum tensor is in congruence with the definition of ADM mass in a gravitation system \cite{Balasubramanian:1999re} and in fact is proportional to the irreducible mass:
\begin{align}\label{eq:irrmq01.12}
\mathcal{Q}_0=GM_{irr}^2,
\end{align}
where the irreducible mass is the final $ADM$ mass state of a Kerr black hole after it has completed its Penrose process. This relation may have broad generality and larger avenues of application, in particular for extremal black hole CFT correspondences where temperature vanishes, but mass is non-zero. 
\vspace{6pt}  


\acknowledgments{\textbf{Acknowledgments:} We thank Vincent Rodgers for enlightening discussions. This work began just before the passing of Sujeev Wickramasekara on December 28th, 2015. We wish to thank Dr. Wickramasekara for all his enlightening/visionary input and thoughts. He is greatly missed and never forgotten!}


\authorcontributions{\textbf{Author Contributions:} L.R. conceived the idea, performed significant calculations and drafted/wrote the article; S.R. performed significant calculation and  assisted in drafting/writing the article.}


\conflictofinterests{\textbf{Conflicts of Interest:} The authors declare no conflict of interest.} 

\section*{\noindent Abbreviations}\vspace{6pt}\noindent 
The following abbreviations are used in this manuscript:\\

\noindent CFT: Conformal Field Theory\\
$AdS$: Anti-de Sitter\\
RW: Robinson and Wilczek\\
ASG: Asymptotic Symmetry Group\\
RN: Reissner-N\"ordstrom\\
NHEK: near horizon extremal Kerr\\
CWG: Conformal Weyl Gravity\\
ADM: Arnowitt, Deser and Misner\\
RW2DA: Robinson and Wilczek two dimensional analog\\
WRSS: Weyl Rescaled Schwarzschild Spacetime\\
EMT: Energy momentum tensor\\
NH: Near Horizon\\
MUBC: modified Unruh Vacuum boundary conditions\\
$Diff$: Diffeomorphism\\
CWGS: Conformal Weyl Gravity solution\\
CWDG: Conformal Weyl Dilaton Gravity

\appendix 







\bibliography{cftgr}
\bibliographystyle{mdpi}


\end{document}